\documentstyle[times,pramana,epsf,floats]{ias}
\begin{document}

\mark{
{Mesoscopic Effects in the Quantum Hall Regime}
{R.~N. Bhatt and Xin Wan}
}
%%%%%%%%%%%%%%%%%%%%%%%%%%%%%%%%%%%%%%%%%%%%%%%%%%%%%%%%%%%%%%%%%%%%%
\title{Mesoscopic Effects in the Quantum Hall Regime}

\author{
R. N. Bhatt and 
Xin Wan\footnote{Present address: National High Magnetic Field Laboratory, 
Florida State University, Tallahassee, FL 32310, USA}
}

\address{Department of Electrical Engineering, Princeton University,
        Princeton, NJ 08544-5263, USA}


\abstract{
We report results of a study of (integer) quantum Hall transitions 
in a single or multiple Landau levels for non-interacting electrons
in disordered two-dimensional systems,
obtained by projecting a tight-binding Hamiltonian to
corresponding magnetic subbands. 
In finite-size systems, we find that mesoscopic effects often dominate, 
leading to apparent
non-universal scaling behaviour in higher Landau levels. This is because
localization length, which grows
exponentially with Landau level index, exceeds the system sizes
amenable to numerical study at present. 
When band mixing between multiple Landau levels is present, mesoscopic
effects cause a crossover from a sequence of quantum Hall transitions for
weak disorder to classical behaviour for strong disorder.
This behaviour may be of relevance to experimentally observed transitions 
between quantum Hall states and the insulating phase at low magnetic
fields.
} 


\maketitle
%%%%%%%%%%%%%%%%%%%%%%%%%%%%%%%%%%%%%%%%%%%%%%%%%%%%%%%%%%%%%%%%%%%%%
\section{Introduction}
\label{sec:introduction}
One of the central ideas used to analyse phase transitions
between different integer quantum Hall states in two-dimensional
electron systems is finite size scaling. 
Transitions between
plateaus in the Hall conductivity are characterized by
a localization length $\xi$ which diverges
at the critical point $s_c$, $\xi(s) \sim |s-s_c|^{-\nu}$.
$s$ can be energy $E$ (in numerical calculations), or magnetic
field $B$ (in experiments). Finite size scaling asserts that
various quantities for a finite system of size L will depend 
on the ratio $L / \xi$. (In experiments, the
relevant size is the dephasing length, which is determined by
temperature $T$ through inelastic scattering mechanisms). Thus,
a system realizes its finite size when $L / \xi \leq 1$, when
singularities associated with the phase transition are cut off
by the finite size. 
This defines the range where finite-size effects become important. 
Therefore, long length scales always pose a challenge for numerical
calculations, in which the system size is limited by the current
computing power. 
Unfortunately, the integer quantum Hall transitions in higher Landau
levels are among such cases. 

To a large extent, the difficulties of identifying the nature of the
quantum Hall transitions in weak magnetic fields come from the same
origin. 
In this case, the mixing of neighbouring Landau levels exacerbates the
situation even further. 
According to the global phase diagram, developed by early theoretical
efforts~\cite{khmelnitskii84,laughlin84,kivelson92},
for integer quantum Hall transitions, the Hall conductance can
only change by $e^2/h$ at the transitions, whether in high magnetic fields or
in low fields.
Early experiments~\cite{jiang93,hughes94,wang94} were consistent with
this expectation
(counting spin degeneracy). However, more recent experiments by Song
{\it et al.}~\cite{song97}, Hilke {\it et al.}~\cite{hilke00},
and Kravchenko {\it et al.}~\cite{kravchenko95}
cast doubts on the global phase diagram, as their data were interpreted
as showing direct transitions from an integer quantum Hall state with
Hall conductance $ne^2 /h$ ($n > 1$) to an insulator. 
Subsequently, some numerical calculations~\cite{sheng99,sheng00}
appeared to corroborate the existence of direct transitions between
quantum Hall liquids with Hall conductance of $ne^2 /h$ ($n > 1$) to
the low field insulator. 
However, it has never become clear whether such ``transitions'' are true
quantum phase transitions or merely the manifestation of consecutive
transitions which are too close to resolve.  

In this paper, we discuss the mesoscopic effects in the quantum Hall
regime, in particular the effects of long length scales on the scaling
behaviour and on the low-field transitions, using a tight-binding
lattice model of two-dimensional non-interacting electrons in a random
potential and a perpendicular magnetic field. 
We describe our model with truncated Hilbert space and various
methods in Sec.~\ref{sec:model} and ~\ref{sec:numbers} respectively. 
Section~\ref{sec:single} contains the discussions on the difference of
the scaling behaviour in Landau levels with increasing index, as a result
of the increasing length scales. 
We then turn to transitions at low magnetic fields in
Sec.~\ref{sec:multiple}, before summarizing our results and conclusions 
in Sec.~\ref{sec:conclusion}

%%%%%%%%%%%%%%%%%%%%%%%%%%%%%%%%%%%%%%%%%%%%%%%%%%%%%%%%%%%%%%%%%%%%%
\section{Model}
\label{sec:model}
We consider a tight-binding Hamiltonian of non-interacting electrons
on a two-dimensional square lattice with nearest neighbour hopping,
subject to a uniform perpendicular magnetic field ${\bf B}$:
\begin{equation}
\label{eq:Hamiltonian}
H = - t \sum_{\langle ij \rangle}{(e^{i \theta_{ij}} c^{\dagger}_{i}
c_{j}
        + h.c.)} + \sum_{i}{\epsilon_{i} c^{\dagger}_{i} c_{i}}. 
\end{equation}
$\epsilon_{i}$ are the on-site random potentials, and the magnitude of the hopping
matrix element $t = 1$ is chosen as the unit of energy for convenience. 
As a consequence of the perpendicular magnetic field ${\bf B}$,
there is a magnetic flux $\phi$ per unit
cell given by 
\begin{equation}
{\phi \over \phi_0} = {e B a_0^2 \over h c} = {1 \over 2 \pi}
\sum_{\Box}
\theta_{ij},
\end{equation} 
where $\phi_0 = hc / e$ is the magnetic flux quantum, and $a_0$ is the
lattice constant.
The random potential $\epsilon_{i}$, are taken to be {\it independent}
random variables, chosen from an identical distribution (so
ensemble averaged quantities are translationally invariant).
We consider two specific distributions:
\begin{enumerate}
\item[(i)] a rectangular box (i.e. uniform) distribution on 
$[-W, W]$, 
\begin{equation}
P(\epsilon) = \left \{ 
\begin{array}{ll}
{1 \over 2W}, & -W \leq \epsilon \leq W, \\
0, & {\rm otherwise,}
\end{array} 
\right .
\end{equation}
\item[(ii)] a Gaussian distribution with standard deviation $\sigma$,
\begin{equation}
\label{gaussian}
P(\epsilon) = {1 \over \sqrt{2 \pi} \sigma } \exp 
\left ( - \frac{\epsilon^2}{2 \sigma^2} \right ).
\end{equation} 
\end{enumerate}

\begin{figure}
\epsfxsize=3.2in
\centerline{ \epsffile{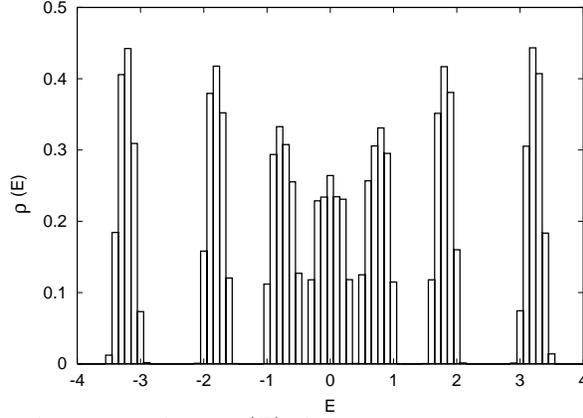} }
\caption{ 
\label{fig:dos}
Density of states $\rho(E)$ of a $14 \times 14$ square lattice 
with magnetic flux $\phi_0 / 7$ per unit cell and $W = 0.5$ 
rectangular box distributed disorder. } 
\end{figure}

The calculation of the energy spectrum can be done for rational
$\phi / \phi_0 = p / q$, where $p$ and $q$ are integers, chosen to
have no common factors.  
The whole spectrum~\cite{hofstadter76} is commonly known as the Hofstadter butterfly,
so named because of the  butterfly-like self-similar pattern
in terms of the parameter
$\phi$.   
For each pair of $p$ and $q$, in the absence of disorder, the original
tight-binding band is split into $q$ magnetic subbands; these are then
broadened by disorder.
Figure~\ref{fig:dos} shows the density of states of a 14 by 14 square
lattice with magnetic flux $\phi_0 / 7$ per unit cell.
The seven magnetic subbands are well separated by energy gaps at
disorder strength $W = 0.5$. 

The Hall conductance of {\em each} of the $q$ magnetic subbands,
when they are separated by 
energy gaps, has been shown to be quantized~\cite{thouless82}.  
The Hall conductance $\sigma_r$ (in units of $e^2 / h$),
when the lowest $r$ subbands are occupied, is related to $p$ and $q$ by
a Diophantine equation~\cite{niu85}, $r = q s_r + p \sigma_r$, where
$s_r$
is an integer chosen such that $|\sigma_r| \leq q / 2$. 
The quantized Hall conductance has a topological origin, which is the
Chern number of a fiber bundle defined by the magnetic Bloch
wavefunctions on the torus, the magnetic Brillouin zone.
For the case of $p = 1$ and  $q > 1$ an odd integer, each of the $(q -
1)$ side subbands carries Hall conductance $e^2 / h$. 
The center subband, on the other hand, has Hall conductance of $-(q - 1)
e^2 / h$, guaranteeing that the total Hall conductance of all the
subbands 
of the tight-binding system is zero.
The continuum model is achieved in the limit $ q \rightarrow \infty $,
where each of the side subbands corresponds to a Landau level, while the
center subband, with no counterpart in the continuum model, is
mapped to infinite energy. For finite $q$, it is expected that the
side subbands near the edges of the spectrum, correspond
closely to the Landau levels in the continuum
model.

Just like in the continuum model\cite{huo92}, we can project the Hamiltonian to a
subspace spanned by the eigenstates in the subbands of interest,
obtaining from diagonalizing the tight-binding Hamiltonian in the
absence of disorder. 
This truncation process is equivalent to the addition of a
pseudopotential that pushes the subbands not of interest to positive or
negative infinite energy. 
The projection can be easily justified in the strong field limit (weak disorder), 
where there are energy gaps separating the subbands. 
On the other hand, in the case of weak magnetic fields (strong disorder),
numerical
calculations show that the center subband widens with 
increasing disorder\cite{yang96,sheng98}
and the negative Hall conductance carried by the central subband
neutralizes the Hall conductance of the opposite sign carried by side
subbands. This has been interpreted as quantum phase transitions from the
quantum Hall liquids (with arbitrary integer filling factors) to
the insulating state\cite{sheng98}.
However, since there is no center subband in the continuum model,
we have chosen to work with a projected Hamiltonian in a truncated
Hilbert space with multiple side subbands only. This has a major advantage
that unphysical effects due to the center subband in the discrete lattice model
are avoided. (We note that the actual experimental system is on a discrete
lattice, but that is on an atomic scale, while the physics we are
concerned with is on the scale of the magnetic length, which is two
orders of magnitude larger).  
We believe the projected
Hamiltonian should keep the essential features of the low magnetic field
physics, at least near the bottom of the spectrum where interactions
with truncated higher Landau levels are weak. 

%%%%%%%%%%%%%%%%%%%%%%%%%%%%%%%%%%%%%%%%%%%%%%%%%%%%%%%%%%%%%%%%%%%%%
\section{Chern Number and Thouless Number}
\label{sec:numbers}
We consider a system of size $L_x \times L_y$ with generalized periodic 
boundary condition
\begin{equation}
\label{eq:boundaryCondition}
t({\bf L}_{x,y}) | m \rangle = e^{i \theta_{x,y}} | m \rangle
\end{equation}
where $t({\bf L}_{x,y})$ is the magnetic translation operator in x or y 
direction. The Hall conductance of an eigenstate can be calculated by
the Kubo formula
as~\cite{huo94}   
\begin{equation}
\label{eq:kuboFormular}
\sigma_{xy} (m; \theta_x, \theta_y) = 
         \frac{i e^2 \hbar}{L_x L_y} \sum_{n \neq m} 
  \frac{\langle m | v_x | n \rangle \langle n | v_y | m \rangle
  - h.c.}
  {(E_n - E_m)^2}.
\end{equation}
Here $| m \rangle$, $| n \rangle$ are eigenstates for the 
boundary condition Eq.~(\ref{eq:boundaryCondition}) and $v_x$, $v_y$
are components of the velocity operator. 
Thouless {\it et al.} showed that the averaged Hall conductance 
can be written as a quantized 
integral~\cite{thouless82}
\begin{equation}
\label{chernNumber}
\sigma_{xy}(m) \equiv {1 \over (2 \pi)^2} \int d \theta_x d \theta_y 
         \sigma_{xy} (m; \theta_x, \theta_y) = C (m) {e^2 \over h}, 
\end{equation}
where $C(m)$ is an integer, known as the Chern number of the
eigenstate.  
States with non-zero Chern number conduct Hall current and can be
identified as conducting states~\cite{arovas88}.
The total Hall conductance $\sigma_{xy}(E_f)$ for a given position of
the Fermi level $E_f$ is obtained by summing over the Hall conductances
of the filled states.  

The longitudinal conductivity $\sigma_{xx}$, related to
Thouless number, may also be used to distinguish
extended states from localized states~\cite{edwards72}. 
Thouless number, which describes the sensitivity of an eigenstate to the
change in boundary conditions, is defined in dimensionless
form~\cite{licciardello75} by the equation: 
\begin{equation}
\label{thouless}
g_L (E) = {\langle \delta E \rangle \over \langle \Delta E \rangle}
\sim {h \over e^2} \sigma_{xx}, 
\end{equation}
where $\langle \delta E \rangle$ is the average shift in the energy
level due to a change of boundary condition from periodic to
anti-periodic, and $\langle \Delta E \rangle = 1 / L^2 \rho_L (E)$ 
the mean energy level separation. In
second-order perturbation theory, the change in eigenenergy
is proportional to the longitudinal conductance~\cite{licciardello75};
the proportionality constant is of the
order of unity, with a precise value that depends on the
exact definition of the energy shift $\langle \delta E \rangle$. 

As the size of the system increases, the Thouless number in the 
localized regime decreases, while at the critical point 
the Thouless number goes to a universal value. In the scaling regime, 
the dimensionless quantity depends on system size
only through the ratio $L / \xi(E)$ : 
\begin{equation}
\label{eq:UniversalFunction}
g_L(E) = \tilde{g} (L / \xi) = g(L^{1 / \nu} |E - E_c|),
\end{equation}
where the localization length $\xi$ diverges at the critical energy
$E_c$, in a power law fashion,   
\begin{equation}
\label{eq:LocalizationLength}
\xi(E) \sim |E - E_c|^{-\nu}.
\end{equation}

We define the width of a Thouless number curve $g_L (E)$ as
\begin{equation}
\label{eq:TNWidth}
W(L) = { 1 \over W_{\rm DOS} (L)}
        \left [ { \int_{-\infty}^{\infty} (E - E_c)^2 g_L (E) dE
        \over \int_{-\infty}^{\infty} g_L (E) dE} \right ]^{1/2} .
\end{equation}
$W(L)$ is normalized by $W_{\rm DOS} (L)$, the width of the
corresponding density-of-states $\rho_L (E)$,
\begin{equation}
\label{eq:DOSWidth}
W_{\rm DOS} (L) =  \left [ { \int_{-\infty}^{\infty} (E - E_c)^2 \rho_L
(E) dE
        \over \int_{-\infty}^{\infty} \rho_L(E) dE} \right ]^{1/2} ,
\end{equation}
which is expected to be only very weakly dependent on $L$. 
Using Eq.~(\ref{eq:LocalizationLength}), the width of a
Thouless number curve can be shown to vary with system size as a power law:
\begin{equation}
\label{eq:TNScaling}
W(L) \sim L^{-1 / \nu}.
\end{equation}

Alternatively, we may compute the area
$A(L)$ under $g_L(E)$, as $ A(L) = \int_{-\infty}^{\infty} g_L (E) dE $.
In the scaling regime, we have
\begin{equation}
\label{eq:areaScaling}
A(L) \sim L^{-1 / \nu}
\end{equation}
with the same exponent $\nu$.
%%%%%%%%%%%%%%%%%%%%%%%%%%%%%%%%%%%%%%%%%%%%%%%%%%%%%%%%%%%%%%%%%%%%%
\section{Quantum Hall Transitions in a Single Subband}
\label{sec:single}
%--------------------------------------------------------------------
\subsection{The lowest subband}
We start with the simplest case - truncation of the Hilbert space to the lowest subband,
which we compare with the numerous
results in the lowest Landau level available to date.
The critical behaviour of the quantum Hall transition in the lowest
Landau level has been studied numerically by various 
groups~\cite{huo92,bhatt91,huckestein95,liudz94,chalker88}. 
Numerical results in different models agree that the localization length 
diverges with critical exponent $\nu \simeq 2.35 \pm 0.1$ in the
vicinity of the critical energy $E_c$ at the band center. 
In particular, Huo and Bhatt found the localization length
critical exponent  $\nu = 2.4 \pm 0.1$, obtained from the finite-size 
scaling of the density of extended states identified by a Chern number
calculation~\cite{huo92}.

\begin{figure}
\epsfxsize=3.2in
\centerline{ \epsffile{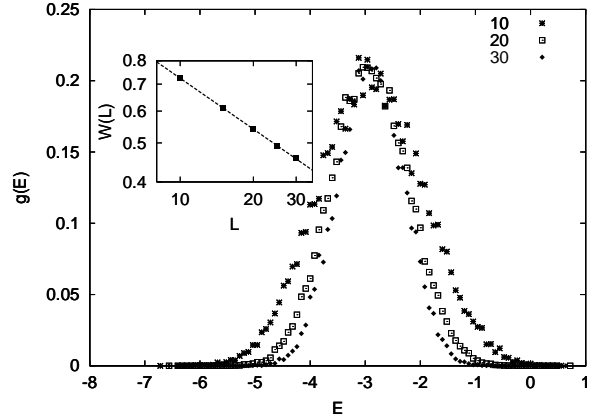} }
\caption{ 
\label{sideband}
Thouless number $g(E)$ of the lowest subband with $W / t = 5$ 
for $10 \times 10$, $20 \times 20$, and $30 \times 30$ lattices. 
The inset shows the normalized width of $g(E)$ as a function 
of size $L$, on a log-log scale. 
} 
\end{figure}

If we choose $\phi / \phi_0 = 1/5$, 
the tight-binding band splits into five magnetic subbands, 
which carry Hall conductance of 
1, 1, -4, 1, 1, in units of $e^2 / h$, respectively.
We separate the
lowest subband carrying Hall conductance $e^2 / h$ from the rest of the
four subbands (carrying total Hall conductance $-e^2 / h$)
by truncating the Hilbert space as described above. 
We studied systems with size $L=10$, 15, 20, 25 and 30, with  
a rectangular box distributed random potential $W / t = 5$. 
We diagonalized between 200 and 1,800 samples with different
random potential configurations from the ensemble, for calculating
average quantities, depending on size. 
Figure~\ref{sideband} shows the Thouless number of the edge subband
for sizes $L=10$, 20 and 30.
We clearly see the shrinkage of the width of the Thouless number as 
the system size increases.
Meanwhile, the peak value of the Thouless number for different sizes
remains fixed around $g_0 \simeq 0.21$, 
which helps us locate the critical energy $E_c = -2.9$. 
The width of Thouless number of different sizes calculated by 
Eq.~(\ref{eq:TNWidth}) (and normalized by the width of the 
density-of-states), is plotted in the inset on a log-log scale. 
Fitting the width to a straight line (a power law), we can extract the
localization length exponent $\nu = 2.4 \pm 0.1$ according to
Eq.~(\ref{eq:TNScaling}), which agrees with the exponent found for the
lowest Landau level in the continuum model.  
The finite-size scaling fit suggests that in the thermodynamic limit
($L \rightarrow \infty$) the Thouless number, 
or the longitudinal conductance $\sigma_{xx}$,
vanishes everywhere except at one single energy $E_c$, 
where the localization length diverges.   
This demonstrates that, under truncation, the edge subband
has the same scaling behaviour as the lowest Landau level.

In order to test the universality of the scaling behaviour, 
we studied systems with various values of magnetic flux per unit cell, 
including $\phi / \phi_0$ = $1/3$, $1/5$, $1/6$, $1/7$, and $1/9$,
choosing either rectangular box distributed or Gaussian distributed 
potential, with disorder strength $W / t$ (or $\sigma / t$) varying from
$1$ 
to $\infty$ ($t \rightarrow 0$).
We calculated the Thouless number of the truncated lowest magnetic
subband
of these systems up to $L = 66$ and averaged over as many as 100,000
random
potential realizations (for $L = 3$).
With these choices of system configurations, we found the localization
length 
critical exponent $\nu = 2.3 \pm 0.2$, 
obtained from the finite-size scaling of the width of the Thouless
number.
The peak value $g_0$ of the Thouless number was found for all different
cases
to be $g_0 = 0.21 \pm 0.02$.
This indicates that $\nu$ and $g_0$ are universal in the scaling regime, 
which includes system sizes as small as $L = 10$ in most cases.
%--------------------------------------------------------------------
\subsection{The second lowest subband}
It has been recognized that the localization length, 
\begin{equation}
\label{LocalizationLength}
\xi = \xi_0 \left \vert {E-E_c \over \Delta E} \right \vert^{-\nu},
\end{equation} 
in higher Landau levels is much larger than that in the lowest Landau
level~\cite{huckestein95,guo}. In the above equation, $\Delta E$ is the width of the
density of states $\rho(E)$. 
$\xi_0$, the $E$-independent prefactor, is distinct for each Landau level, and is of
order of the lattice constant in the lowest Landau level.
The localization length in higher Landau levels grows rapidly with
increasing Landau level number (exponentially as its square in
simple approximations), raising great difficulties in numerical
calculation of the critical exponent in higher Landau levels. 
This difficulty has been reflected, in particular from the observation
of non-universal (larger-than-expected) exponent, in early 
works\cite{huckestein95,guo,aoki85,ando85,mieck90,mieck93}.  
The universal scaling behaviour can, therefore, only be restored in much
larger systems, unless the localization length can be reduced. 

A potential with a non-zero correlation length is known to
be able to reduce the localization length substantially, to a
numerically acceptable value.
Calculations for the second lowest Landau level suggest that the value 
of the localization length exponent $\nu$ depends on the correlation
length $l$ of the random potential~\cite{guo,mieck93}. 
Mieck showed that $\nu$ decreases from 6.2 for $l = 0$ to 2.3 for $l = 4 
l_c$, where $l_c = (\hbar c/ eB)^{1/2}$ is the magnetic
length~\cite{mieck90,mieck93}.  
Huckestein also obtained for a correlation length $l = l_c$ the same
scaling behaviour as in the lowest Landau level~\cite{huckestein92}. 
We found very similar behaviour in the truncated second lowest
magnetic subband, 
which, again, shows the correspondence between our lattice model
with the truncated Hilbert space and the continuum model.

In our lattice model, a correlated potential is generated as follows. 
First, generate an uncorrelated random potential $\epsilon_i^0$ with
the desired distribution. 
We then introduce correlation between random potential on different
sites by defining
\begin{equation}
\label{eq:correlation}
\epsilon_i = \frac {\sum_j \epsilon_j^0 e^{-d_{ij}^2 / l^2} }
        {\sum_j e^{-d_{ij}^2 / l^2} },
\end{equation}
where $d_{ij}$ is the distance between sites $i$ and $j$.
The correlation function between the two sites
\begin{equation}
\langle \epsilon_i \epsilon_j \rangle \sim e^{-d_{ij}^2 / l^2},
\end{equation}
with correlation length $l$.
It should be emphasized that our correlated potentials have Gaussian
correlation; to get an exponential correlation, an appropriate
modification of Eq.~(\ref{eq:correlation}) is necessary. 

\begin{figure}
\epsfxsize=3.2in
\centerline{ \epsffile{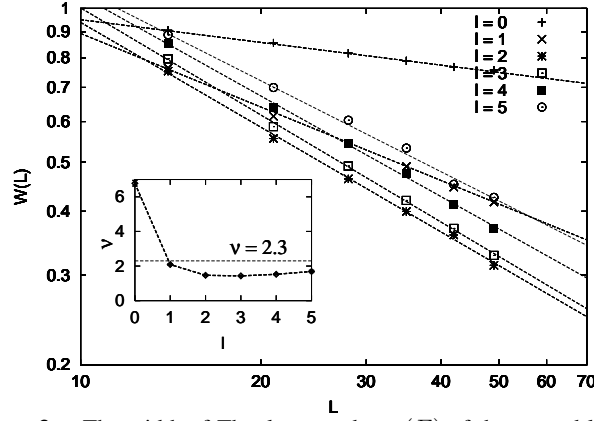} }
\caption{ 
\label{width7.2}
The width of Thouless number $g(E)$ of the second lowest subband as a
function of size $L$ on a log-log scale. 
The inset shows the localization length critical exponent $\nu$
calculated from the slopes of the different lines in main figure
for different correlation lengths $l$. 
} 
\end{figure}

In this section, we study the Thouless number of the (truncated) second
lowest subband, of a seven-subband system ($\phi / \phi_0 = 1/7$), 
with correlated potential generated using a rectangular
box distribution for $\epsilon_i$.
After projection we set $t =0$ (infinite disorder limit) to
reduce finite size effects.
For sizes $L$ = 14 to 49, we diagonalized 150 - 1,800 samples, 
depending on $L$. 
We calculated the Thouless number $g(E)$ of the subband with various
correlation lengths. 
Figure~\ref{width7.2} summarizes the width of the $g(E)$
normalized by the width of the density of states.
On the double logarithmic plot, the width versus size follows straight
lines apparently consistent with the scaling hypothesis
Eq.~(\ref{eq:TNScaling}).
However, the localization length critical exponent $\nu$ extracted from
the slope of the curves is not universal. 
The inset of Fig.~\ref{width7.2} shows that $\nu$ decreases from 6.8
with $l = 0$ to $1.4$ with $l = 3$, and then increases slowly to $1.7$
with $l = 5$. 
This curve is very similar to the one obtained by Guo~\cite{guo} in the
study of the critical behaviour in the first Landau level in the
continuum model. 
The curve crosses the expected universal exponent $\nu = 2.3$ for
$l \simeq 1$. 

Figure~\ref{width7.2} also shows that $W(L)$ is minimized with $l = 2$,
suggesting that the localization length is reduced with 
increasing correlation length $l$ of the correlated potential, 
at least for $l \leq 2$ . 
However, with $l = 2$, $\nu = 1.5$ is much smaller than the expected
$\nu = 2.3$.  
This suggests that the random correlated potential introduces an
irrelevant length scale $\zeta$, which depends on the correlation length 
$l$, so that 
\begin{equation}
g_L(E) = \tilde{g} \left ({L \over \xi}, {L \over \zeta} \right),
\end{equation}
Only when $L, \xi \gg \zeta$ can we restore the universal  
behaviour of quantum Hall transitions. 
Due to this additional length scale $\zeta$, Thouless number, in
general, does not have either the universal functional form or the
universal peak value $g_0$ in finite systems. 

We might expect the irrelevant length scale $\zeta$ to be less important as
the localization length $\xi$ increases and dominates when the band
center ($E_c = 0$) is approached.
On the other hand, the irrelevant variable clearly affects our results in the tails
where $\xi$ is small. 
The width of the Thouless number defined by Eq.~(\ref{eq:TNWidth}), 
however, depends strongly on the scaling behaviour of
the tail, since the value of $g(E)$ in the tails is amplified by
multiplying $(E - E_c)^2$ in the numerator. 
Therefore, the localization length exponent obtained from the
finite-size scaling of the area $A(L)$ of the Thouless number
may be more reliable (which includes the finite-size effect of
the peak conductance as well). 

\begin{figure}
\epsfxsize=3.2in
\centerline{ \epsffile{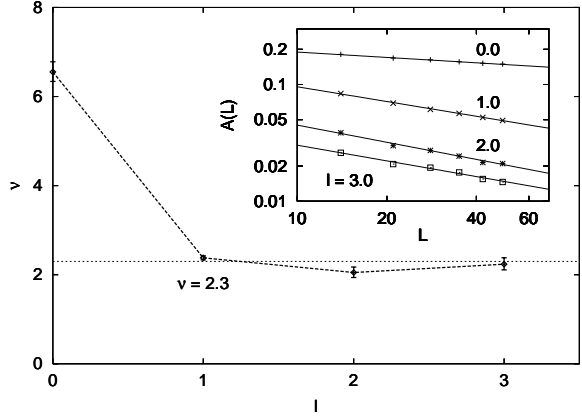} }
\caption{ 
\label{area7.2}
The localization length critical exponent $\nu$ in the second lowest
subband,
calculated for different correlation length $l$ of the random potential,
from the
scaling of the area $A(L)$ with size, $L$. 
The inset shows the area $A(L)$ of Thouless number $g(E)$ of the second
lowest
subband as a function of size $L$ on a log-log scale. 
} 
\end{figure}

Figure~\ref{area7.2} inset shows the area of Thouless number $g(E)$ of
the
second lowest subband as a function of size $L$. 
On the log-log plot, the area $A(L)$ versus $L$ follows straight lines with
different slopes.
The numerical value of the localization length critical exponent $\nu$
calculated based on Eq.~(\ref{eq:areaScaling}) is summarized in
Fig.~\ref{area7.2}. 
The universal value $\nu = 2.2 \pm 0.2$ is restored for $l \geq 1$. 
These values of $l$ are consistent with the observation in
Fig.~\ref{width7.2} that $W(L)$ is minimized with $l = 2$. 
$\nu = 6.5$ with $l = 0$ is, again, due to the finite size effects as a
result of the large localization length. 

%--------------------------------------------------------------------
\subsection{The third lowest subband}
\begin{figure}
\epsfxsize=3.2in
\centerline{ \epsffile{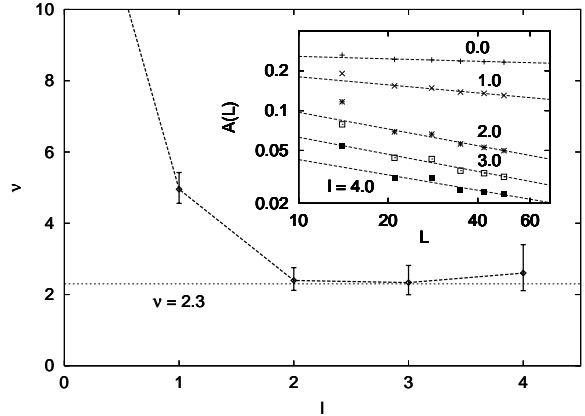} }
\caption{ 
\label{area7.3}
The localization length critical exponent $\nu$ calculated at different
correlation length $l$. 
The inset shows the area of Thouless number $g(E)$ of the third lowest
subband as a function of size $L$ on a log-log scale. 
} 
\end{figure}

In even higher subbands (Landau levels), the localization length $\xi$
is expected to be even larger than $\xi$ in the second subband, 
with both uncorrelated and correlated potential. 
We confirmed, in our calculations, that the effects of Gaussian
correlated potential on the scaling behaviour of the third lowest (side)
subband of the seven-subband system agree qualitatively with that of
the second lowest subband.  
Once again, to minimize the effect of the irrelevant length scale
$\zeta$, we plot the area $A(L)$ of the Thouless number curve versus the
system size $L$ in the inset of Fig.~\ref{area7.3}. 
For $L \geq 21$, $A(L)$ follows straight lines on double logarithmic
plot.
Figure~\ref{area7.3} summarizes the localization length critical
exponent
$\nu$ calculated according to Eq.~(\ref{eq:areaScaling}). 
$\nu$ decreases from $16$ with $l = 0$ to $\nu \simeq 2.3$ with $l \geq
2$. 
The $\nu$ versus $l$ curve is very similar to the curve in
Fig.~\ref{area7.2}, except that, in the second subband, $\nu$ is smaller
with $l = 0$, and drops to $\nu \simeq 2.3$ with $l = 1$. 
This is consistent with the fact that the localization length $\xi$ is
smaller in lower subband. 

Therefore, due to rapidly increasing localization length in higher
Landau levels and the consequent mesoscopic effects, finite size scaling
of the area $A$ under the Thouless number curves seems to be more reliable
than the width of the Thouless number curve.
A universal exponent $\nu \simeq 2.3$ can be observed for correlated
potential with certain correlation length in finite size samples.
It is worth emphasizing that the universality, including localization
length exponent and critical conductance, should be observed in large
enough samples. These are beyond our current computing capabilities,
but should become more accessible as computers get more powerful. 

%%%%%%%%%%%%%%%%%%%%%%%%%%%%%%%%%%%%%%%%%%%%%%%%%%%%%%%%%%%%%%%%%%%%%
\section{Quantum Hall Transitions in Multiple Subbands}
\label{sec:multiple}

In this section, we briefly present our results for multiple
subbands. More details can be found elsewhere~\cite{wan00,wan01b}. 
In a lattice with $\phi / \phi_0 = 1/7$, we projected the Hamiltonian to
the lowest three subbands, carrying a total Hall conductance of
$3e^2/h$. 
We found that, with increasing disorder, Thouless number $g(n)$,
calculated as a function of filling factor $n$, evolves
from three peaks separated by gaps at integer fillings to a smooth curve
with no dips.  
Meanwhile, the plateaus of Hall conductance $\sigma_{xy}$ at integer
fillings, characteristic of quantum Hall transitions, disappears. 

\begin{figure}
\epsfxsize=3.2in
\centerline{ \epsffile{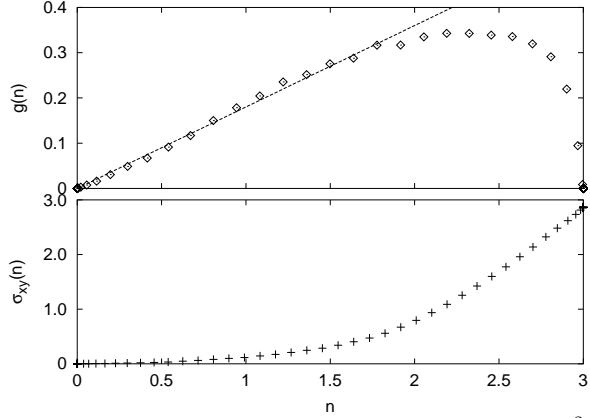} }
\caption{ 
\label{threebands}
Thouless number and Hall conductance (in units of $e^2/h$) in the
truncated, three lowest subbands with $W / t = 4.5$.
} 
\end{figure}

Figure~\ref{threebands} shows $g(n)$ and $\sigma_{xy}(n)$ as a function
of filling $n$, with disorder strength $W / t$ = 4.5, which corresponds to
the strong disorder (weak field) limit. 
We calculated $g(n)$ in a $28 \times 28$ lattice, and $\sigma_{xy}(n)$ in a
$14 \times 14$ lattice. 
The dip in $g(n)$ near the upper end of the spectrum
is affected by the truncation of higher subbands. 
However, at the lower end, we believe our data shows the correct
low-field physics, since the energy levels in the subbands not included
in the truncated model
interact only weakly with the low-lying energy levels. 
For $n \leq 1.5$, we find that $g(n)$ can be fit by a straight line,
while the Hall conductance $\sigma_{xy}$ is much smaller than the diagonal
conductance, and is essentially zero. 
Such observations are generically available for systems of different
magnetic flux and number of subbands. 

We attribute our results to the ``classical'' behaviour of the
two-dimensional electron gas. 
The linear dependence of the longitudinal conductance on electron
concentration is consistent with the classical Drude conductivity
$\sigma_{xx} \sim n$. 
This
result seems to suggest that for finite sizes, due to the presence of
long length scales, quantum interference effects can be
cut off at strong disorder, or equivalently at weak magnetic field. This could
explain the experimental phase diagram~\cite{hilke00}. 
The concurrent vanishing $\sigma_{xy}$ (lower panel of
Figure~\ref{threebands}) supports this scenario too. For further
details, the reader is referred to Ref.~\cite{wan01b}.

%%%%%%%%%%%%%%%%%%%%%%%%%%%%%%%%%%%%%%%%%%%%%%%%%%%%%%%%%%%%%%%%%%%%%
\section{Conclusions}
\label{sec:conclusion}
In summary, we have studied a tight-binding lattice model for non-interacting
electrons, using projection to enable truncation of the Hilbert
space to different combinations of the magnetic subbands.
In a single magnetic (side) subband, we obtained, through finite-size
scaling of Thouless number, the critical exponent $\nu$ of the 
localization length $\xi \sim |E - E_c|^{-\nu}$, as $\nu = 2.4 \pm 0.1$,
in good agreement with previous numerical results in continuum lowest 
Landau level model.
The localization length critical exponent is universal, regardless of
the 
strength and distribution of the random potential or magnetic field.
In higher Landau levels, however, the rapidly increasing length scales
can alter the scaling behaviour in finite size systems, 
a manifestation of mesoscopic
effects in the quantum Hall regime. 
With the introduction of correlated potential, we demonstrated that the
localization length can be reduced, leading to universal scaling
behaviour of appropriately chosen quantities. 

In a truncated system involving multiple magnetic side subbands 
(corresponding to multiple Landau levels), we found that the Hall
conductance 
$\sigma_{xy}$ of finite lattices evolves from a series of well-defined
plateaus at $ne^2/h$ for various integer Landau level filling to a
single sharp rise at the upper end of the energy spectrum.
At the same time, the Thouless number of the (finite) system become
linear in filling factor $n$. 
The linear dependence is interpreted as a crossover to the classical 
(unquantized) Hall response, as the system size becomes comparable to, 
or smaller than, the very (exponentially) large localization lengths of 
the electronic eigenstates in this limit. 
While the sizes studied here are considerably smaller than those 
corresponding to experimental samples, our findings raise the question
of 
the proper interpretation of the experimentally observed low field 
transition, and the possibility of the influence of a quantum-classical 
crossover in experiments.

%%%%%%%%%%%%%%%%%%%%%%%%%%%%%%%%%%%%%%%%%%%%%%%%%%%%%%%%%%%%%%%%%%%%%
\section{Acknowledgments}

It is a pleasure to participate in the commemorative symposium (and its
proceedings) in honor of Professor Narendra Kumar, who has made many
significant contributions to the field of disordered and mesoscopic systems.
This research was funded by NSF DMR-9809483.

%%%%%%%%%%%%%%%%%%%%%%%%%%%%%%%%%%%%%%%%%%%%%%%%%%%%%%%%%%%%%%%%%%%%%

%%%%%%%%%%%%%%%%%%%%%%%%%%%%%%%%%%%%%%%%%%%%%%%%%%%%%%%%%%%%%%%%%%%%%
\end{document}